# HALO: Report and Predicted Response Times

Matthew Swisher
The Ohio State University
Columbus, USA
swisher.80@osu.edu

*Abstract*—HALO: Heterogeneity-Aware Load Balancing is a paper that proposes a class of heterogeneity-aware Load Balancers (LBs) for cluster systems. LBs that are heterogeneity-aware are able to detect when servers differ in speeds and in number of cores. Response times for heterogeneous systems are calculated and presented.

## I. INTRODUCTION

HALO: Heterogeneity-Aware Load Balancing is a paper written by Anshul Gandhi, Xi Zhang, and Naman Mittal, all of who are affiliated with Stony Brook University. The document presents three of the most popularly used scalable Load Balancers (LBs). The paper then introduces the improved HALO version of each of the three LBs, and provides numerical, simulation, and implementation results of HALO LBs. The performance of the HALO LBs is compared to the performance of the three original LBs that the HALO LBs were built on.

Performance is compared when heterogeneity is due to differences in server speed, and number of cores. Server speed differs when certain components, such as the processor, differ. Core count between server groups may vary when VMs are rented from cloud service providers. The paper is able to provide simulation results for cases when heterogeneity is due to server speed. Only numerical results are provided for instances where heterogeneity is due to number of cores. This is due to the difficulty in obtaining a closed-form expression for optimal load split.

In this paper an attempt to reproduce the optimal response times, T*, is presented by using equations that are found in HALO: Heterogeneity-Aware Load Balancing [1]. The equations are applied to *k* simple M/G/1/PS models.

## II. HALO OVERVIEW

Load Balancers (LBs) are important for distributed systems [1]. They disperse traffic, or load, among clusters [1]. Load is measured by request rate over throughput [1]. Scalable LBs are needed to handle the huge number of requests encountered by some online service providers [1]. Scalable LBs do not split loads optimally due to their heterogeneity-unaware designs [1]. Heterogeneity occurs when servers differ in processor speed, and when the number of cores in each server differs [1]. Server speeds are measured by requests per second [1].

A few of the most popular scalable LBs currently used in industry are Randomized (RND), Round-Robin (RR), and Power-of-D LBs [1]. RND LBs randomly select a server to send requests to, RR LBs select which server to send requests to by rotating through each of the servers successively, and POD LBs, similar to RND LBs, randomly select a server to send a request to [1]. RND and POD LBs differ in that POD LBs randomly select a server from a small number D of the total available servers [1]. Two types of POD routing are used in the HALO research, and are referred to as Jsq_POD and Base_POD. Jsq_POD sets D equal to the sum of $k_i$ with *i* from 1 to *n*, where *n* is the number of groups in a heterogeneous cluster of servers [1]. Base_POD sets D equal to 2 [1].

HALO LBs were developed by starting with the three previously mentioned LBs, and modifying their request distribution, so that optimal performance in heterogeneous environments could be obtained [1]. The request distribution was modified by using queuing-theory analysis [1]. The goal was to figure out how to distribute requests to obtain the lowest average response time possible [1].

Optimal load split distributes load among servers in a way that results in the best possible average response time or T*[1]. Optimal load split is found by using equations they derived by optimizing the equation used to figure mean response time [1]. Concerning the HALO LBs, optimal load split approaches probabilistic load split as load increases [1]. Probabilistic load split is the probability that a given server will receive a request, and illustrates the situation where load is split proportionally [1].

The HALO algorithms always perform better than all the other algorithms except for a few load values when compared to Base_RR and Jsq_POD [1]. HALO LBs provide the most improvement when heterogeneity is due to difference in server speeds [1]. HALO LBs are scalable [1]. HALO LBs are not the first heterogeneity-aware LBs, but are the first class of scalable, heterogeneity-aware LBs [1]. There is no increase in overhead when HALO LBs are used [1].

## III. RESPONSE TIME PREDICTIONS

In order to predict T* for a heterogeneous cluster of *n* M/G/1/PS servers using RND LB with speeds $\mu_i$ (*i* = 1, 2,…*n*) and total request rate $\lambda$ the following equation is used:

$$T^* = \frac{2\sum_{i=1}^{n}\sum_{j=i+1}^{n}(k_i k_j \sqrt{\mu_i \mu_j}) - \sum_{i=1}^{n}(k_i \mu_i \sum_{j=1,j\neq i}^{n} k_j) + \lambda \sum_{i=1}^{n} k_i}{\lambda \left(\sum_{i=1}^{n}(k_i \mu_i) - \lambda\right)}$$

In order to predict mean response time T for a network of M/G/1/PS servers we use the equation:

$$T = \sum_{i=1}^{n} p_i T_i = \sum_{i=1}^{n} \frac{p_i}{\mu_i - \lambda p_i / k_i}$$

For $n = 2$, $\mu_1 = 2$, $\mu_2 = 1$, $k_1 = 1$, $k_2 = 2$, and $\lambda = \{0.8, 1.6, 2.4, 3.2\}$ req/s the following mean and optimal response times are obtained.

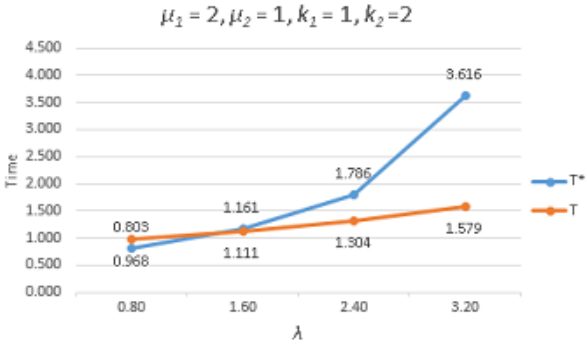

For $n = 2$, $\mu_1 = 1.5$, $\mu_2 = 1$, $k_1 = k_2 = 2$, and $\lambda = \{1, 2, 3, 4\}$ req/s the following mean and optimal response times are obtained.

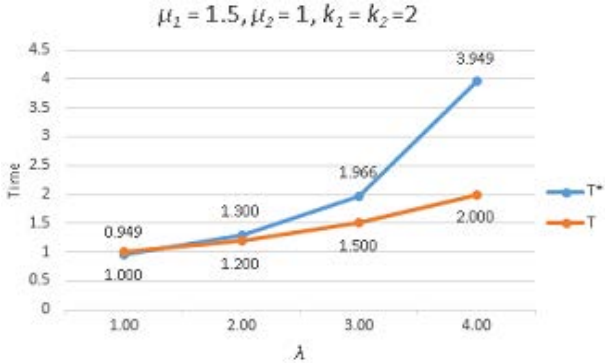

Results achieved for T* in both graphs are not as expected. For each graph T* only performs better than T when $\lambda = 0.8$ in the first graph, and when $\lambda = 1$ in the second graph. These values of lambda result in a load of 0.2. Load is calculated with the following equation:

$$\rho = \lambda / \left( \sum_{i=1}^{n} k_i \cdot \mu_i \right)$$

It is unclear as to why values for T* are higher than the values for T. These results do not agree with the results presented in the HALO research paper. A reason for discrepancies is most likely due to misunderstanding what models should be used when attempting to reproduce the results achieved by comparing response times. All equations used during the attempt to reproduce results found in HALO: Heterogeneity-Aware Load Balancing are obtained from that report [1].

IV. RELATED WORK

Adaptive, Model-Driven Autoscaling for Cloud Applications is a paper on research that went into building s cloud service they call Dependable Compute Cloud (DC2). Information is presented that shows that applications that have a dynamic workload and are on cloud servers need to be resized so they cost-effectively meet performance requirements. DC2 does this automatically, and dynamically by providing an application-agnostic cloud offering [2].

DC2 automatically and proactively scales infrastructure. It uses a queuing-network model to approximate multi-tier cloud applications. The queuing-network model is aided by Kalman filtering technique to infer applications unobservable parameters [2].

Unlike the research done to develop the HALO algorithms, homogeneous servers and perfect load balancing are used in the research that went into creating DC2. Both instances of research aim to improve performance and minimize resource costs [2].

Another paper that is similar in research is AutoScale: Dynamic, Robust Capacity Management for Multi-Tier data Centers. A dynamic capacity management policy called AutoScale is introduced. AutoScale reduces the number of servers needed in data centers. Servers are left on while idle. Considering that servers are on average only busy 10-30% of the time, this can greatly reduce wasted energy by adding or removing servers as needed. In order to handle request rate bursts, AutoScale provides the perfect amount of spare capacity thanks to the capacity inference algorithm. The appropriate capacity can be given blindly in response to changes in request rate, server efficiency, or request size [3].

This technology would do well with LBs that use HALO routing algorithms. For servers that differ in speed, HALO algorithms send all of the load to the faster server when request rates are low. AutoScale would be able to shut down the slower, idling server when faster servers are handling all of the work.

A third research paper relates to the HALO LB algorithms by its focus on scaling and how it can be used to better define the term elasticity. This paper is titled, 'Elasticity in Cloud Computing: What It Is and What It Is Not', and it defines elasticity as, "the degree to which a system is able to adapt to workload changes by provisioning and deprovisioning resources in an autonomic manner, such that at each point in time the available resources match the current demand as closely as possible." [4]. The HALO algorithms are the first scalable, heterogeneity-aware LBs, and by the definition given of elasticity, the HALO algorithms add elasticity to LBs.

For the same reason as the previously mention research paper, the paper 'ShuttleDB: Database-Aware Elasticity in the Cloud' relates to the HALO research by focusing on efficient scalability. Elasticity is also discussed in this document [5].

REFERENCES

[1] A. Gandhi, X. Zhang and N. Mittal, "HALO: Heterogeneity-Aware Load Balancing," *Modeling, Analysis and Simulation of*

*Computer and Telecommunication Systems (MASCOTS), 2015 IEEE 23rd International Symposium on*, pp. 242-251, 2015.

[2] A. Gandhi, P. Dube, A. Karve, A. Kochut, and L. Zhang, "Adaptive, Model-driven Autoscaling for Cloud Applications," *Proceedings of the 11th International Conference on Autonomic Computing (ICAC)*, Philadelphia, PA, 2014.

[3] A. Gandhi, M. Harchol-Balter, R. Raghunathan and M. Kozuch, "AutoScale: Dynamic, Robust Capacity Management for Multi-Tier Data Centers," *Transactions on Computer Systems*, vol.30, 2012.

[4] N. Herbst, S. Kounev and R. Reussner, "Elasticity in Cloud Computing: What It Is, and What It Is Not," *10th International Conference on Autonomic Computing (ICAC)*, San Jose, CA, 2013.

[5] S. Barker, Y. Chi, H. Hacigümüs, P. Shenoy and E. Cecchet, "ShuttleDB: Database-Aware Elasticity in the Cloud," *Proceedings of the 11th International Conference on Autonomic Computing (ICAC)*, Philadelphia, PA, 2014.